\documentclass{elsart}

\usepackage{graphicx}

\usepackage{bm}
\usepackage{amsfonts}
\usepackage{amssymb}
\usepackage{latexsym}
\usepackage{dcolumn}

\newcommand{\commentout}[1]{}
\newcommand{\nwc}{\newcommand}
\nwc{\rer}{r_{\eta,\rho}}
\nwc{\rinf}{r_{\eta,\infty}}

\nwc{\xvec}{{\vec{\bx}}}
\nwc{\kvec}{{\vec{\bk}}}
\newcommand{\lt}{\left}
\newcommand{\rt}{\right}

\newcommand{\ks}{\omega}

\newcommand{\bx}{\mathbf x}
\nwc{\mm}{\mathbf m}
\newcommand{\br}{\mathbf r}

\nwc{\bS}{\mathbf S}
\newcommand{\bp}{\mathbf p}
\newcommand{\by}{\mathbf y}
\nwc{\bX}{\mathbf X}
\nwc{\bY}{\mathbf Y}
\nwc{\bh}{\mathbf h}

\newcommand{\bw}{\mathbf w}

\newcommand{\bH}{\mathbf H}
\nwc{\bQ}{\mathbf Q}
\nwc{\bI}{\mathbf I}

\nwc{\nwt}{\newtheorem}
\nwt{assumption}{Assumption}

\nwc{\bal}{\begin{align}}
\nwc{\beq}{\begin{equation}}
\nwc{\ben}{\begin{equation*}}
\nwc{\bea}{\begin{eqnarray}}
\nwc{\beqa}{\begin{eqnarray}}
\nwc{\bean}{\begin{eqnarray*}}
\nwc{\beqn}{\begin{eqnarray*}}
\nwc{\beqast}{\begin{eqnarray*}}

\nwc{\eal}{\end{align}}
\nwc{\eeq}{\end{equation}}
\nwc{\een}{\end{equation*}}
\nwc{\eea}{\end{eqnarray}}
\nwc{\eeqa}{\end{eqnarray}}
\nwc{\eean}{\end{eqnarray*}}
\nwc{\eeqn}{\end{eqnarray*}}
\nwc{\eeqast}{\end{eqnarray*}}

\nwc{\tx}{\tilde{\bx}}
\nwc{\tp}{\tilde{\bp}}
\nwc{\tr}{\tilde{\br}}
\nwc{\tw}{\tilde{\bw}}
\nwc{\ep}{\varepsilon}
\nwc{\ept}{\epsilon}
\nwc{\vrho}{\varrho}
\nwc{\orho}{\bar\varrho}
\nwc{\ou}{\bar u}
\nwc{\vpsi}{\varpsi}
\nwc{\lamb}{\lambda}
\nwc{\wep}{W^\ep}

\nwc{\nn}{\nonumber}
\nwc{\mf}{\mathbf}
\nwc{\mb}{\mathbf}
\nwc{\ml}{\mathcal}

\nwc{\IA}{\mathbb{A}} 
\nwc{\IB}{\mathbb{B}}
\nwc{\IC}{\mathbb{C}} 
\nwc{\ID}{\mathbb{D}} 
\nwc{\IM}{\mathbb{M}} 
\nwc{\IP}{\mathbb{P}} 
\nwc{\II}{\mathbb{I}} 
\nwc{\IE}{\mathbb{E}} 
\nwc{\IF}{\mathbb{F}} 
\nwc{\IG}{\mathbb{G}} 
\nwc{\IN}{\mathbb{N}} 
\nwc{\IQ}{\mathbb{Q}} 
\nwc{\IR}{\mathbb{R}} 
\nwc{\IT}{\mathbb{T}} 
\nwc{\IZ}{\mathbb{Z}} 

\nwc{\cE}{{\ml E}}
\nwc{\cI}{{\ml I}}
\nwc{\cP}{{\ml P}}
\nwc{\cL}{{\ml L}}
\nwc{\cR}{{\ml R}}
\nwc{\cV}{{\ml V}}
\nwc{\cW}{{\ml W}}
\nwc{\cT}{{\ml T}}
\nwc{\crV}{{\ml V}_{(\delta,\rho)}}
\nwc{\cC}{{\ml C}}
\nwc{\cA}{{\ml A}}
\nwc{\cS}{{\ml S}}
\nwc{\cK}{{\ml K}}
\nwc{\cB}{{\ml B}}
\nwc{\cD}{{\ml D}}
\nwc{\cF}{{\ml F}}
\nwc{\cM}{{\ml M}}
\nwc{\cN}{{\ml N}}
\nwc{\cG}{{\ml G}}
\nwc{\cH}{{\ml H}}
\nwc{\bk}{{\mb k}}
\nwc{\cQ}{{\ml Q}}
\nwc{\cO}{{\ml O}}
\nwc{\cJ}{{\ml J}}
\nwc{\sir}{{\sf SIR}}
\nwc{\snr}{{\sf SNR}}
\nwc{\sinr}{{\sf SINR}}

\nwc{\mint}{{\int\cdot\int}}

\begin{document}
\begin{frontmatter}\title{Time Reversal Communication in Rayleigh-Fading Broadcast Channels with Pinholes}
\author{Albert C. Fannjiang}
 \ead{
  cafannjiang@ucdavis.edu}
   \ead[url]{http://www.math.ucdavis.edu/\~\,fannjian}
 \thanks{
 I thank the American Institute of Mathematics and the organizers of the Workshop
 ``Time-Reversal Communications in Richly Scattering Environments'', October 18-22, 2004, for a stimulating
 meeting which motivated  the present work.
 This research is partially supported by
 U.S. National Science Foundation grant DMS 0306659.
}
 \address[al]{
Department of Mathematics,
University of California, Davis 95616-8633}

\begin{abstract}
The paper presents an analysis of the time reversal 
in independent-multipath Rayleigh-fading channels with $N$
inputs (transmitters) and $M$ outputs (receivers).
 The main issues addressed 
are the condition of statistical stability, the rate of
information transfer and the effect of pinholes.
The stability condition is proved to be 
 $MC\ll N_{\rm eff}B$ for broadband channels and
 $M\ll N_{\rm eff}$ for narrowband channels where $C$ is the symbol rate,
 $B$ is the bandwidth and 
$N_{\rm eff}$ is the {\em effective} number (maybe less than 1) of
transmitters. It is shown
that when the number of screens, $n-1$, is relatively low
compared to the logarithm of numbers of pinholes $N_{\rm eff}$  is given by 
 the {\em harmonic} (or {\em inverse}) {\em sum} of the number of transmitters and
the numbers of pinholes at all screens. 
 The novel idea of the effective number
of time reversal array (TRA) elements is introduced to  derive the stability condition and
estimate  the channel capacity
in the presence of multi-screen  pinholes. 
The information rate, under the  constraints of the noise power $\nu$ per unit 
frequency and the average total power $P$, attains the supremum $P/\nu$ in the regime $M\wedge N_{\rm eff}\gg P/(\nu B)$.  In particular,
when $N_{\rm eff}\gg M\gg P/(B\nu)$ the optimal information
rate can be achieved with statistically stable, sharply focused 
signals. 
\end{abstract}

\end{frontmatter}

\maketitle
\section{Introduction}
\begin{figure}
\begin{center}
\includegraphics[width=8cm, totalheight=5cm]{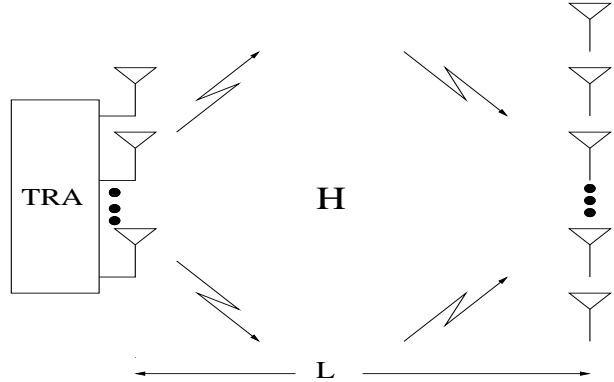}
\end{center}
\caption{MIMO Broadcast Channel
 }
 \end{figure}

Time reversal (TR) of waves has received
great attention in recent years and been extensively studied 
for electromagnetic \cite{BL},  \cite{LF} as well as acoustic propagation (see \cite{Fink} and the references therein). A striking effect of time reversal
in randomly inhomogeneous media is the superresolution
of refocal signals \cite{BPZ}, \cite{tire-phys}  which implies low probability
of intercept and holds high potential in technological
applications such as 
communications \cite{DTF}, \cite{RJD},  \cite{EK}, \cite{KK},\cite{KKP}.

An issue prior to superresolution, however, is
statistical  stability, namely the question:
 How many antennas and how much bandwidth 
does one need 
to achieve self-averaging in TR so that the received signals are nearly
deterministic, independent of the channel statistics?
 In this paper
we answer this question for independent-multipath Rayleigh fading
channels,  with
multiple inputs and multiple outputs (MIMO),  commonly used in wireless communication literature, see, e.g. \cite{Paul}.
We also introduce the novel idea of effective number of transmitters
to  analyze the effect of multi-screen pinholes 
on stability and capacity.

In the MIMO-TR communication scheme as studied  in  \cite{DTF}, \cite{pulsa-pnas}, the $M$
well-separated receivers  first send a pilot signal to
the $N$-element time reversal array (TRA) which then uses the time-reversed
version of the received signals to modulate the data symbols and retransmit them
back to the
receivers.
 One of the main results obtained here is that
the time reversal process is statistically stable when
\bea
\label{stab}
MC&\ll N_{\rm eff}B,&\quad\hbox{for broadband channels}\\
M&\ll N_{\rm eff},&\quad\hbox{for narrowband channels}
\label{stab2}
\eea
 where $C(\leq 2B)$ is the symbol rate,
 $B$ is the bandwidth and 
$N_{\rm eff}$ is the {\em effective} number of
transmitters (maybe less than one). In the presence of $(n-1)$-screen pinholes, 
we show that the  effective number of
transmitters is asymptotically 
 the harmonic sum of
the number of transmitters and the number
of pinholes of every screen when all  these numbers
are greater  than  $2^n$. 

The LHS of (\ref{stab}) is the number of degrees of freedom
per unit time
in the constellation of input data-streams while
the RHS of (\ref{stab}) is roughly 
the number of degrees of freedom per unit time in the channel state information (CSI)
received by TRA from the pilot signals.  The latter
has to be larger than the former in order to reverse 
the random scrambling by the channel and achieve
deterministic outputs. The stability condition $N\gg 1$
for narrow-band channels
or $B\gg \beta_c$ (the coherence bandwidth) for
broadband channels, when $M$ is  small  and the pinholes 
are absent,  have been previously discussed in
\cite{BPZ}, \cite{DTF1}, \cite{DTF2}, \cite{DTF}, \cite{LF}.

In Section~4 and 5.2,  we analyze the information
rate of the TR broadcast channel in the presence of noise.
We show that
the optimal  information rate $R\sim P/\nu$, under the power
and noise constraints,  can be achieved in the regime 
$M\wedge N_{\rm eff}\gg P/({\nu B})$
 where $\nu$ is the magnitude
of noise per unit frequency and 
$P$ the average total power input.  In particular,
when $N_{\rm eff}\gg M\gg P/(B\nu)$ the optimal information
rate can be achieved with statistically stable, sharply focused 
signals. 

\section{TR-MIMO communication}
 First let us review the MIMO-TR communication scheme
 as described in \cite{pulsa-pnas} which is an example
 of broadcast channel \cite{Paul}.

 The $M$
receivers  located at $\by_j, j=1,...,M$ first send a pilot signal $
\int e^{i {\ks t}}g(B^{-1}(\ks-\ks_0))d\ks\delta(\bx-\by_j)$ to
the $N$-element TRA located at $\bx_i, i=1,...,N$ which then uses the time-reversed
version of the received signal $\int e^{i {\ks t}}g(B^{-1}(\ks-\omega_0)) H(\by_j, \bx_i;\ks)d\ks$ to encode a stream of symbols and retransmit them
back to the
receivers. Here $H$ is the transfer function of
 the propagation channel at the frequency $\ks$
 from point $\by$ to $\bx$ and  $g^2(\ks)$ is the power
density at $\ks$.  Let $\bH(\ks)=[H_{ij}(\ks)],
 H_{ij}(\ks)=H(\bx_i,\by_j;\ks),$ be the transfer matrix between
 the transmitters and receivers. The reciprocity implies
 that $H(\by_j,\bx_i;\omega)=H_{ij}(\omega)$ and 
 $\bH^*(\ks)=\bH(-\ks)$ where $*$ stands for complex conjugation.   Let us assume
that $g$ is a smooth and rapidly decaying
function such as the Gaussian function.
  Naturally  
 the relative bandwidth $B/\ks_0$ is less than unity so that
 $\omega_0\gg 1$ 
 if $B\gg 1$. 
 In this paper we will assume $B/\omega_0\ll 1$
 to simplify the frequency coherence structure below
 (Section~3). 
 We have chosen the time unit such that
the speed of propagation is one and the wavenumber equals
the frequency. 

 The
  signal vector $\bS=(S_j)$ arriving at the  receivers with delay $L+t$ is then given by \cite{pulsa-pnas} (see also \cite{BPZ}, \cite{DLF})
  \beqa
{S_j(t)}
&= &\sum_{l=1}^W\sum_{i=1}^M m_i(\tau_l)\int  e^{-i{\ks}(t-\tau_l)}g(\frac{\ks-\ks_0}{B})
\sum_{k=1}^NH_{jk}(\omega)H^*_{ik}(\ks) d\ks
\label{mr}
\eeqa
where $m_j(\tau_l), l=1,...,W\leq\infty$ is a stream of symbols intended for the $j$-th
receiver transmitted at times $\tau_l=l\tau,\tau>0 $.  
In vector notation, we have $\bS
=\sum_{l=1}^W \int e^{-i\ks(t-\tau_l)}g(B^{-1}(\ks-\ks_0))\bH\bH^\dagger(\ks)\mm(\tau_l)
  d\ks$ where
 $\bH^\dagger $ is the conjugate transpose of $\bH$ and
 $\mm(\tau_l)=(m_j(\tau_l))$.
Let us note that while all the TRA-elements are coordinated
and 
synchronized the receivers do not know
the channel
 and can not coordinate in decoding the total signal vector
received. As a consequence, the multi-user interference
 arises and can  be a serious impedance to communications.
An advantage of the time reversal scheme is the possibility to use
the (statistical) stability property to achieve
the following asymptotic
\[\hspace{-.5cm}
\int  e^{-i{\ks}(t-\tau_l)}g(\frac{\ks-\ks_0}{B})\sum_{k=1}^NH_{jk}(\omega)H^*_{ik}(\ks) d\ks
\sim B\delta_{ij}e^{-i\omega_0(t-\tau_l)}\cF^{-1}[g](B(\tau_l-t))
\]
so that $S_j(t)\sim B\sum_{l=1}^Wm_j(\tau_l) e^{-\omega_0(t-\tau_l)}\cF^{-1}[g](B(\tau_l-t))$ and each receiver receives the input symbols with
little  interference.  Here and below $\cF^{-1}$ stands for the inverse Fourier transform. 

\section{Statistical stability}
 One of
the main goals of the present note is to characterize  the stability regime
for the  independent-multipath Rayleigh fading channel in which $H_{ij}(\omega)$ are 
 independent $\cC\cN(0,\sigma)$, the zero-mean, variance-$\sigma$ circularly symmetric complex-Gaussian random variables
 and $\big\{H_{ij}(\omega)\big\}_{i,j,\omega}$ are a {\em jointly}  Gaussian process. The independent-multipath Rayleigh fading  is an idealized model for richly scattering
 environment,  after proper normalization,
 when the spacings  within the transmitters
 and  receivers are larger than 
 the coherence length $\ell_c$ of the channel.
In general, the coherence length  is inversely proportional to
 the angular spread \cite{Paul} and sometimes
 can be computed explicitly in terms of physical properties
 of the channel \cite{pulsa-pnas}. 
For diffuse waves the coherence length is known to be on the scale of
wavelength \cite{Sh}, \cite{SS}. 

We set the variance $\sigma=1/(N\vee M)$ so that
the average input  power is no less than 
the average output power. The value of $\sigma$ would not
change the conditions of statistical stability but will affect
the discussion of information transfer in the next section.

 Let us calculate
 the mean and the variance of the signals with respect to
 the ensemble of the channel.  Let  $\IE$  denote
 the channel ensemble average. For simplicity, we assume that $|m_i(\tau_l)|=\mu, $$
 \forall i,l$. 
 By the Gaussian rule for
 the calculation of moments we have
  \bea
 \IE\bS=B N\sigma\mm \sum_{l=1}^We^{-i\ks_0(t-\tau_l)}
 \cF^{-1}[g](B(t-\tau_l)). \label{a.42}
 \eea
  Let $\tau\geq (2B)^{-1}$ so that the  summation in $\IE \bS$ is $B$-uniformly bounded as $W\to\infty$. 
 
The statistical stability of
 the signals can be measured by  the normalized variance  of the signals at the receiver $j$
 \[
 \cV_j(\tau_n)=\frac{V_j(\tau_n)}{|\IE S_j|^2(\tau_n)},
 \,\, V_j(\tau_n)\equiv \IE |S_j|^2(\tau_n)-|\IE S_j(\tau_n)|^2,
 \]
 $\forall j, n$
 and the time-reversed signals are stable when $\cV_j(\tau_n)\to 0, \forall j,n$. Note that $\cV_j^{-1}(\tau_l)$ is exactly
 the signal-to-interference ratio ($\sir$) at receiver $j$.
 
 Let $\beta_c$ be  the coherence bandwidth of the channel such that
 \[
 \IE\big[H_{ij}(\omega) H^*_{i'j'}(\omega')\big]
 \approx \sigma f(\omega_0, \frac{\omega-\omega'}{\beta_c}) \delta_{ii'}\delta_{jj'}
 \]
 where $f(\omega_0, \cdot)$ is a  continuous,  rapidly decaying function and $f(\omega_0, 0)=1$ (see \cite{2f-whn},
 \cite{pulsa-pnas}  for a
 rigorous example). Here we have used the fact that the relative bandwidth $B/\omega_0$ is small so that $f$ is independent
 of the precise value of the frequency. Below we shall suppress 
 the argument $\omega_0$ in $f$. 
The coherence bandwidth $\beta_c$ is  inversely proportional
 to the delay spread  and hence 
 the delay-spread-bandwidth product (DSB) is roughly $B\beta_c^{-1}$ \cite{2f-whn}, \cite{pulsa-pnas}, \cite{Paul}.  
 In the diffusion approximation $\beta_c$ is given 
 by the Thouless frequency $D_BL^{-2}$ where
 $D_B$ is the Boltzmann diffusion constant, equal to the energy transport velocity times the transport mean free path, and $L$ the distance
 of propagation  \cite{LT}, \cite{Sheng}. 
 
The {\em broadband, frequency-selective} (BBFS) channel
is naturally defined as having a large DSB, i.e. $B\beta^{-1}_c\gg 1$.
Since $B<\ks_0$, $\ks\in [\ks_0-B/2, \ks_0+B/2]$ and $-\ks$ are separated by more than $\beta_c$ in a BBFS channel. On the
other hand, $B\ll \beta_c$ corresponds to the  {\em  narrow-band, frequency-flat} (NBFF) channel. For convenience in  the subsequent analysis,
we shall think of the NBFF channel as the limit $\beta_c\to\infty$ and the BBFS channel as the limit 
$\beta_c\to 0$ while  $\omega_0, B$ are fixed.
In either case, we have
 \bea
 V_j(t)\label{obs}
 &\approx&N\sigma^2\sum_{i=1}^M\sum_{l, l'=1}^Wm_i(\tau_l)m_i^*(\tau_{l'})
e^{i\omega_0(\tau_l-\tau_{l'})}\\
 &&\times \int d\omega d\omega' e^{-i(\omega-\omega')(t-\tau_l)} e^{i\omega'(\tau_l-\tau_{l'})} g(\frac{\omega}{B})g^*(\frac{\omega'}{B})|f|^2(\frac{\omega-\omega'}{\beta_c}).\nn
  \eea
  
Consider the NBFF channels first. We  obtain by passing
to the limit  $\beta_c\to\infty$ 
  \bean
 V_j(t)\approx N\sigma^2 B^2|f|^2(0)\sum_{i=1}^M\Big|\sum_{l=1}^Wm_i(\tau_l)e^{i\ks_0\tau_l}\cF^{-1}[g](B(t-\tau_l))\Big|^2.
  \eean
In view of (\ref{a.42})  the stability condition $N\gg M$ for NBFF channels then follows easily. 
On the other hand, the BBFS channels 
($\beta_c\to 0$) yields
  \bea
 \label{1.3}V_j(t)&\approx&N\sigma^2 \sum_{i=1}^M\sum_{l, l'=1}^Wm_i(\tau_l)m_i^*(\tau_{l'})
e^{i\omega_0(\tau_l-\tau_{l'})}\nn\\
 &&\times \int d\omega'' d\omega' e^{-i\omega''(t-\tau_l)} e^{i\omega'(\tau_l-\tau_{l'})} g(\frac{\omega'}{B})g^*(\frac{\omega'}{B})|f|^2(\frac{\omega''}{\beta_c})\nn\\
 &\approx&N\sigma^2 B\beta_c\sum_{i=1}^M\sum_{l=1}^Wm_i(\tau_l)\cF^{-1}\big[{|f|^2}\big](\beta_c(\tau l-t))\nn\\
 &&\times\sum_{l'=1}^Wm_i^*(\tau_{l'})e^{i\ks_0\tau(l-{l'})}\cF^{-1}[|g|^2](B\tau(l-l')).
  \eea
Several observations are in order.  First,
due to $\tau\geq (2 B)^{-1}$  the summation over $l'$ in (\ref{1.3})
is convergent as $W\to \infty$ uniformly in $B$.
Second, due to the term  $\cF^{-1}\big[{|f|^2}\big](\beta_c(\tau l-t))$, there
are effectively $
C\beta_c^{-1}$ terms in the summation over $l$ where $C=\tau^{-1}$ is the number of symbols per unit time in {each}
 data-stream. As a result, the variance 
 $
V_j{\sim}  N\sigma^2 BMC\mu^2$ is independent of $\beta_c$. 
 It then follows  that
 $\cV_j\to 0$ if and only if
 $NB\gg MC$ for BBFS channels. The transition to the
 condition $N\gg M$ for NBFF channels takes place
 when $B\sim C$, i.e. $\tau\sim B^{-1}$.
 
 The stability condition can be interpreted as follows: $NB$ is the number of degrees of freedom in the CSI  collected 
 at the TRA per unit time;  $MC$ is the number
 of degrees of freedom in the ensemble of messages per unit time; 
 the stability condition $NB\gg MC$ says 
 that in order to recover the input  messages, independent
 of the channel realization, and thus reverse the random scrambling
by the channel,  the former must be much larger
 than the latter. In light of this interpretation, the stability condition
 derived above appears to be sharp.

A detailed,  rigorous analysis 
of the MIMO-TR 
channel modeled 
by a stochastic Schr\"odinger equation, in the parabolic
approximation of scalar waves,  with
a random potential is given in \cite{pulsa-pnas}.

 \section{Rate of information transfer}
 In this section we discuss the information rate for a
memoryless channel  which is constructed out of
the time-invariant channel model analyzed in
Section~3. The temporal dependence is introduced by drawing an independent realization from the  Rayleigh-fading ensemble of  
 transfer matrices  after each use of the channel, i.e. after
 each delay spread (or two if the time for channel estimation
 is included).
 This is obviously  an idealization  but widely used
 in communications literature
 \cite{Tel}, \cite{FG}. The coherence time of the resulting ergodic channel
 model 
is then much longer than one delay spread. 
 We assume as in standard practice that  in addition to the
 random channel fluctuations
 additive-white-Gaussian-noise (AWGN)  is
 present at  each receiver, that the input signal vector is multivariate Gaussian
 and that the channel, the noise and the input signal
 are mutually independent. 
 
 For the Rayleigh fading
  channel prior to adding noise, 
   each frequency component 
   of the time reversed signal $S_j$ in  (\ref{mr})
  \bean
  \label{freq}
&&\sum_{i=1}^M\sum_{k=1}^Nm_i(\tau_l)  g(\frac{\ks-\ks_0}{B})
H_{jk}(\omega)H^*_{ik}(\ks)\\
&=&\underbrace{\sum_{k=1}^Nm_i(\tau_l)  g(\frac{\ks-\ks_0}{B})
H_{jk}(\omega)H^*_{jk}(\ks)}_{\hbox{$N$-degree central  $\chi^2$ r.v.}}+
\underbrace{\sum_{ i\neq j}\sum_{k=1}^Nm_i(\tau_l)  g(\frac{\ks-\ks_0}{B})
H_{jk}(\omega)H^*_{ik}(\ks)}_{\hbox{$N(M-1)$ i.i.d. zero-mean   r.v.s}}\nn
\eean
is a sum of 
a central $\chi^2$ random variable with 
  $N$ degrees of freedom and $N(M-1)$ i.i.d. mean-zero random variables. This is due to the assumption that different
  entries of the transfer matrix are mutually independent  zero-mean Gaussian 
  random variables. Therefore, for $N\gg 1$ the interference
  statistic is approximately Gaussian, by the Central Limit Theorem. More generally, after synthesizing all the available  frequencies, 
 the interference statistic becomes approximately Gaussian if $NB\beta_c^{-1}\gg 1$ which is
  always the case for the BBFS channels. In a BBFS (resp. NBFF) channel,
  $NB\beta_c^{-1}$  (resp. $N$) is the number of independent
  subchannels from TRA to each receiver. 
  
   Moreover, each frequency component of $S_j$ has the mean
  \bea
 \IE\Big[ \sum_{i=1}^M\sum_{k=1}^Nm_i(\tau_l)  g(\frac{\ks-\ks_0}{B})
H_{jk}(\omega)H^*_{ik}(\ks)\Big]
=N\sigma g(\frac{\ks-\ks_0}{B}) m_j(\tau_l).\label{inout}
\eea
which exhibits the simple input-output relation: The $\omega$-component of  the input signal
for the $j$-th receiver is $m_jg(\omega)$ and
the received signal component is $N\sigma m_jg(\ks)$ corrupted by the noise
and interference which for $N\gg 1$ is approximately 
Gaussian. Since the $M$ receivers operate independently of
one another, the total time-reversal broadcast channel
consists of $M$ independent subchannels in parallel each of which
has the above input-output relation. Thus the total information rate  is  the sum of those of
the $M$ subchannels from
  TRA to individual receivers. And, in view of the simple  input-output relation,  each subchannel can be viewed as a single-input-single-output (SISO)
  linear filter channel corrupted by (approximately) Gaussian noise/interference for which 
  Shannon's theorem is applicable.  
  
   According to Shannon's theorem \cite{CT} the ergodic capacity (in nats per unit time and frequency) of a SISO linear filter
   channel  is 
  $\ln{(1+\sinr)}$ where $\sinr$, the signal-to-interference-and-noise ratio at each receiver,  is given
  by  the harmonic sum of the $\sir$, the signal-to-interference ratio and $\snr$, the signal-to-noise ratio,  i.e.
  $\sinr=(\sir^{-1}+\snr^{-1})^{-1}$.  For extension of Shannon's result
 to the MIMO setting, see \cite{FG},
  \cite{Tel}.
  
Analogous to the NBFF channels in Section~3, $\sir(\omega)=\cV_j^{-1}\sim N/M$, independent of $\mu$ and $\omega$.
 Let $\nu$ be the noise power, per unit frequency,  at each receiver. Suppose the average transmission  power is constrained to 
  $P$ and all
  the transmit and receive  antennas are identical. 
  
 Since the value of $\sigma$ would
 affect $\snr$ (but not $\sir$) we discuss the two
 cases $N\geq M$ and $N<M$ separately.

{\bf Case 1: $N\geq M$.} In this case, $\sigma=N^{-1}$ and 
   in  view of (\ref{inout}),  $\snr(\ks)=\mu^2/\nu$
   where $\mu=|m_j|$ can be related to the total power constraint $P$ as
  $
   \mu^2M\sim{P}/B
   $ since the average  input power per unit frequency is
   \[
   \sum_{k=1}^N\sum_{i=1}^M |m_i(\tau_l)|^2 |g|^2(B^{-1}(\omega-\omega_0))
   \IE\big|H_{ik}(\omega)\big|^2\sim MN\sigma \mu^2=M\mu^2.
   \]
 Thus 
  $
   \snr(\ks)\sim {P}/({\nu BM}).
   $
   Therefore the total channel capacity (in nats per unit time) is roughly given by
   \beq
   \label{cap2}
 BM\ln{\Big[1+ \frac{1}{M}\Big(\frac{1}{N}+\frac{\nu B}{P}
   \Big)^{-1}\Big]}.
   \eeq  
   Now we ask the question: What is the maximal
  rate at which  a TRA, with fixed number of elements $N$, fixed  average total power
  $P$ and fixed noise level (per frequency) $\nu$,  can transfer information
  if there is no limitation to the number of receivers $M$  and the bandwidth $B$?

Expression   (\ref{cap2}) can be optimized at
the limit $M\gg P/({\nu B})$ to yield the optimal
information rate of $ P/\nu$ which 
  is linearly  proportional to the power. We see that  the simplest strategy for optimizing the information rate of a given TRA
  under the the power and noise constraints
  is to enlarge the bandwidth $B$ as much as possible.
  And if we can satisfy  $N\gg M\gg P/({\nu B})$ then we can 
  achieve stability as well as  the optimal information rate.  
    
   Consider the thermal noise power $\nu= k_B T$ where
 $k_B$ is the Boltzmann's constant and $T$ the temperature.
 Then the above result implies that the energy cost per nat is $ P/R\sim k_B T$
 which is consistent with the classical result of minimum energy
 $k_B T$ requirement for transmitting one nat information at
 temperature $T$ \cite{Pei}, \cite{Lev}.

 {\bf Case 2. $N\leq M$.}  
  In this case, $\sigma=M^{-1}$ and
  (\ref{inout}) implies that 
 $\snr\sim N^2\mu^2/(M^2\nu)$ where $\mu$ is related
 to $P$ by $\mu^2=P/(NB)$. Hence $\snr\sim NP/(M^2\nu B)$. With $\sir\sim N/M$ and Shannon's theorem, the channel
 capacity is roughly 
 \beq
 BM\ln{\Big(1+\frac{N}{M}\big(1+\frac{M B\nu}{P}\big)^{-1}\Big)}
 \eeq
 which achieves the optimal rate  $P/\nu$ in the regime $N=M\gg P/(B\nu).$
 On the other hand, for $M\ll P/(B\nu)$, the information rate
 becomes $BM \ln{(1+N/M)}\leq BN$ which is much smaller than
 $ P/\nu$.

Therefore we conclude that under the power and noise constraints  the  condition for the optimal information rate $P/\nu$ 
 is $N\geq M \gg P/(B\nu)$, which can be achieved by sufficiently large bandwidth,   whereas the additional condition
 $N\gg M$, which, sufficient for the Gaussian approximation to
 the interference statistic,  would also guarantee stability.

Before ending this section, let us compare the capacity 
in the conventional, non-TR MIMO channel as  calculated  in 
\cite{FG}, \cite{Tel}, \cite{Mou}, \cite{SM}. Consider the non-TR single-user channel with the M transmit antennas (on the right of Fig. 1) which
have no channel knowledge and the $N (\geq M)$ receive antennas (on the left of
Fig. 1) as the single user which has perfect CSI. This is, of course,
the reciprocal case of the TR broadcast channel.  In this case,
$\snr\sim P/(MB \nu)$ and 
it is shown in \cite{FG} and \cite{Tel}  
that the ergodic capacity of the single-user narrowband
Rayleigh-fading  channel  scales like $ BM\ln{\snr}$ at high $\snr$ which  can be recovered from  (\ref{cap2}) by
imposing the additional constraint  $M\leq P/(\nu B)\leq N$. 
And as we learn from the discussion of Case 1 above, this
is {\em not}  the regime for achieving  the optimal  information rate $P/\nu$. 

 The same results as discussed in this section are  obtained for the parabolic Markovian channel model in \cite{pulsa-pnas}. 
 \section{Pinhole effect}
 \label{pin}
 \begin{figure}
\begin{center}
\includegraphics[angle=360,width=8cm, totalheight=4cm]{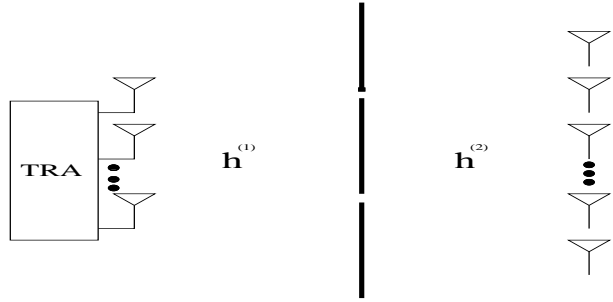}
\end{center}
\caption{Single-screen pinholes
 }
 \end{figure}

 Pinholes are degenerate channels  that can occur 
 in a wide family of channels, outdoor as well as indoor, see
 Fig. 2 and 3. While preserving the co-channel decorrelation,  pinholes have been shown to severely  limit
 the degrees of freedom and reduce the channel capacity \cite{CFV}, \cite{GP}, \cite{CFG}. In this section, we introduce
 the notion of effective number of TRA elements to
  analyze 
 the multi-screen  pinhole effect on  TR in Rayleigh fading.
 
 Let us begin with  the simplest case of single-screen pinholes
 as illustrated in Fig. 2. Let $\bh^{(1)}(\omega)$ be the $N\times K$ transfer matrix  from the TRA to the pinholes and $\bh^{(2)}(\omega)$
 the $K\times M$ transfer matrix from the pinhole
 to the $M$ receivers at frequency $\omega$. The combined channel can be
 described by $\bH(\omega)=\bh^{(2)}(\omega)\bh^{(1)}(\omega)=[\sum_{k=1}^Kh^{(2)}_{ik}(\omega) h^{(1)}_{kj}(\omega)]$
 in which $ h^{(1)}_{kj}(\omega)$ and   $h^{(2)}_{ij}(\omega)$ are assumed
 to be  independent $\cC\cN(0, \sigma_1)$ and $\cC\cN(0,\sigma_2)$,
 respectively, and $\big\{ h^{(1)}_{ij}(\omega), h^{(2)}_{ij}(\omega)\big\}_{i,j,\omega},$ 
 are jointly Gaussian processes. To prevent the average  input power
 from being less than the average  output power we set 
$\IE |H_{ij}|^2=K\sigma_1\sigma_2= (N\vee M)^{-1}, \forall i,j$. Note that the entries of $\bH$ are in general {\em not}
independent r.v.s. 

 As before we assume  the frequency coherence structure
 \beq
 \label{band2}
 \IE\big[h^{(k)}_{ij}(\omega) h^{(k)*}_{i'j'}(\omega')\big]
 \approx \sigma_k f( \frac{\omega-\omega'}{\beta_c}) \delta_{ii'}\delta_{jj'},\quad\forall k
 \eeq
 where, for simplicity,  $f $ and $\beta_c$ are taken to be independent of the screens. Straightforward calculations with the Gaussian rule show that
 the mean signal is
 \[
 \IE\big[S_j(t)\big]=BNK\sigma_1\sigma_2
 \sum_{l=1}^Wm_j(\tau_l)\cF^{-1}[g](B(\tau_l-t))
 \]
 and its variance is
 \bea
\hspace{-1cm}&&V_j(t)=\sigma^2_1\sigma^2_2 NK\sum_{l,l'=1}^W e^{i\omega_0(\tau_l-\tau_{l'})}
\int d\omega d\omega' e^{-i\omega(t-\tau_l)}e^{i\omega'(t-\tau_{l'})} g(\frac{\omega}{B}) g^*(\frac{\omega'}{B})|f|^2(\frac{\omega-\omega'}{\beta_c}) \nn\\
\hspace{-1cm}&&\times
\Big(m_j(\tau_l)m_j^*(\tau_{l'})+ N\sum_{i=1}^M m_i(\tau_l)m_i^*(\tau_{l'})+K|f|^2(\frac{\omega-\omega'}{\beta_c}) \sum_{i=1}^M m_i(\tau_l)m_i^*(\tau_{l'})\Big)\label{9}
\eea
In view of the observations following eq. (\ref{1.3}) we 
have the estimate
 $V_j(t)\sim B^2KN(MN+MK+1)\sigma^2_1\sigma_2^2|\mu|^2$
 for the NBFF channels and 
$V_j(t)\sim BC KN(MN+MK+1)\sigma^2_1\sigma_2^2|\mu|^2$ for
the BBFS channels. As in (\ref{1.3}) the variance
does not depend on the coherence bandwidth $\beta_c$. Therefore we obtain 
 the normalized variance of the signal to the leading order  
 ($N,K\gg 1$) 
 \[
 \cV_j\approx \lt\{\begin{array}{ll}
  {M}\big(N^{-1}+K^{-1}\big),&
 \hbox{for the NBFF channels}\\
 {MC}B^{-1}\big(N^{-1}+K^{-1}\big),&\hbox{for the BBFS channels}.
\end{array}.\rt.
\]
The result suggests the notion of 
 {\em effective} number of TRA-elements given by
$
N_{\rm eff}=NK\big(N+K\big)^{-1},
$
namely  the harmonic sum of $N$ and $K$, so that
$\cV_j\approx MC B^{-1}N_{\rm eff}^{-1}$ for the BBFS channels
and $\cV_j\approx MN_{\rm eff}^{-1}$ for the NBFF channels.
For $N,K\gg 1$ the number of statistically independent paths is  roughly $N_{\rm eff}\times M$.

The previous case {\em without }
 pinholes corresponds to the limiting case $K\gg N$. 
For a fixed  $K$, however,  the previous benefit of stability with large number of TRA elements  
 ($N\gg 1$) disappears.  The multiple antennas in TRA 
 are essentially screened out by the pinholes and
 the effective number  of TRA-elements becomes $K$.
 
 \begin{figure}
\begin{center}
\includegraphics[width=10cm, totalheight=5cm]{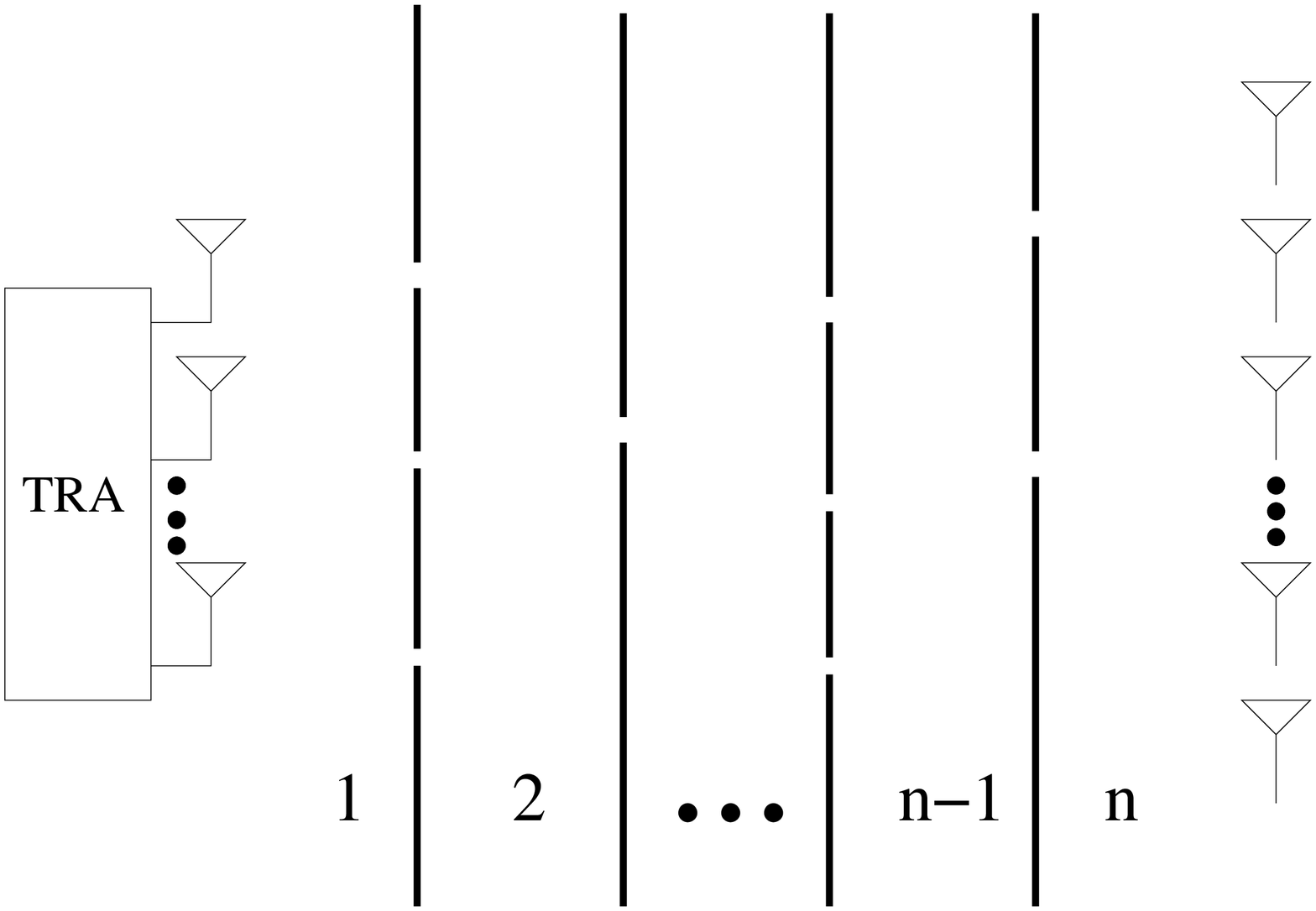}
\end{center}
\caption{Multi-screen pinholes
 }
 \end{figure}

 \subsection{Multi-screen pinholes}
 The same analysis can be applied to 
 channels with $(n-1)$ screens of pinholes such as illustrated in Fig. 3.
 Let $K_k, k=1,..n-1$ be the number of $k$-th screen pinholes.   
 Let $\bh^{(k)}$ be the transfer matrix for the $k$-th stage channel whose entries are  independent $\cC\cN(0, \sigma_k)$   and let the transfer matrices of different
 stages be mutually  independent. Again, in order for
 the average input power to be no less than the average output
 power we  set 
 \beq
 \label{power}
 \IE|H_{ij}|^2=K_1\cdots K_{n-1}\sigma_1\cdots \sigma_n=(N\vee M)^{-1}.
 \eeq
 The condition of statistical stability, however, is independent of the values of $\sigma_k, k=1,...,n$.
 
\begin{figure}
\begin{center}
\includegraphics[width=6.5cm, totalheight=3cm]{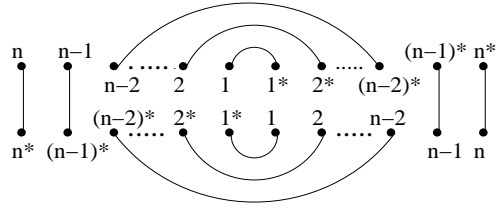}
\end{center}
\caption{Separable diagram: $*$ means complex conjugation;
the top indices are unprimed and
the bottom indices are  primed. 
 }
 \label{graph}
 \end{figure}

 As noted previously the the normalized variance does not depend
on $\beta_c$ and its order of magnitude is determined solely
by the same-frequency moments which will be the focus
of the subsequent calculation.  The calculation of the mean
is straightforward: $\IE (\bH\bH^\dagger\mm)_j=NK_1\cdots  K_{n-1}\sigma_1\cdots\sigma_n m_j$. Let us analyze the second moment of entry $a$
\bea
\label{gauss}\nn
\lefteqn{\IE \Big(\bH\bH^\dagger\mm\Big)_a\Big(\bH\bH^\dagger\mm\Big)_a^*}\\
\nn&=&\IE\Big\{\sum_{i_1,\cdots i_n\atop j_2,\cdots  j_{n+1}}
h_{a i_n}^{(n)}h_{i_n, i_{n-1}}^{(n-1)}\cdots h^{(2)}_{i_3, i_2}h^{(1)}_{i_2, i_1}h^{(1)*}_{j_2, i_1}h^{(2)*}_{j_3, j_2} \cdots h^{(n-1)*}_{j_n, j_{n-1}}
h^{(n)*}_{j_{n+1}, j_n}m_{j_{n+1}}\\
&&\times\sum_{i'_1,\cdots i'_n\atop j'_2,\cdots  j'_{n+1}}
h_{a i'_n}^{(n)*}h_{i'_n, i'_{n-1}}^{(n-1)*}\cdots h^{(2)*}_{i'_3, i'_2}h^{(1)*}_{i'_2, i'_1}h^{(1)}_{j'_2, i'_1}h^{(2)}_{j'_3, j'_2}\cdots h^{(n-1)}_{j'_n, j'_{n-1}}
h^{(n)}_{j'_{n+1}, j_n}m^*_{j'_{n+1}}\Big\}.\nn
\eea
According to the Gaussian rule for computing moments,
the above expression can be represented by $2^n$ diagrams of
$4n$ vertexes and $2n$ edges. We distinguish two types
of edges: 
 the {\em arcs}, connecting (un)primed indices to (un)primed indices,
 and 
the {\em  ladders}, connecting unprimed indices to  primed indices,
see Fig. \ref{graph}.

 When a new screen of pinholes, represented  by $\bh^{(n+1)}$,
is added, the number of diagrams is doubled: one half of them
contain the ladders connecting  $h^{(n+1)}_{ai_{n+1}}$
to $h^{(n+1)*}_{ai'_{n+1}}$ and
$h^{(n+1)*}_{j_{n+2}, j_{n+1}}$ to 
$h^{(n+1)}_{j'_{n+2}, j'_{n+1}}$ while the other half
contain the arcs connecting $h^{(n+1)}_{ai_{n+1}}$
to $h^{(n+1)*}_{j_{n+2}, j_{n+1}}$ and $h^{(n+1)*}_{ai'_{n+1}}$
to $h^{(n+1)}_{j'_{n+2}, j'_{n+1}}$. 
Straightforward calculation with (\ref{band2})  yields the following rule: A new pair of arcs add
 to diagrams with outermost arcs the $K_{n}^2$ (multiplicative) factor and diagrams with outermost ladders the $K_{n}/M$ factor; on
 the other hand,  a new pair of ladders add to diagrams with outermost
 ladders the $K_{n}^2$ factor and diagrams with
 outermost arcs the $K_{n}M$ factor. 
 
 That is,
 the diagrams that correspond to the highest power in
 $K_1, K_2,\cdots$, have the least number of
  edge-type  alternating. Hence for $K_1,\cdots, K_{n-1}\gg 2^n\gg N$
   the leading order term in the variance
  corresponds to the diagram with all ladders and
  is of order $K_1^2\cdots K_{n-1}^2 NM$  while
  the square of the mean corresponds to the diagram with all arcs and is of order $K_1^2\cdots K_{n-1}^2N^2$. 
  The stability condition thus remains the same  as
  in the case without pinholes.

  Let us consider the more interesting regime
 in which  $N,  K_1,..,K_{n-1}\gg 2^n$. 
We claim that to the leading order the normalized variance  of the signal  is given by
$\cV_j\approx MCB^{-1}N_{\rm eff}^{-1}$ where
the effective number of TRA-element $N_{\rm eff}$ is given by
\[
N_{\rm eff}=\Big(N^{-1}+N^{-1}_{\rm p}\Big)^{-1},\quad N_{\rm p}=\Big(\sum_{j=1}^{n-1} K_j^{-1}\Big)^{-1};
\]
namely  the  harmonic sum 
of $N, K_1, \cdots, K_{n-1}$.  
We sketch the proof here. 
The leading order terms in the  variance  after expectation correspond to the
{\em separable} diagrams in which the arcs are nested and are flanked  by
the ladders, Fig. \ref{graph}. Except for the diagram with all ladders, the separable
diagrams all have the innermost arcs connecting
$h^{(1)}_{i_2, i_1}$ to $h^{(1)*}_{j_2, i_1}$ and
$h^{(1)*}_{i'_2, i'_1}$ to $h^{(1)}_{j'_2, i'_1}$, which
give rise to the factors $N^2$ (an extra $N$ than otherwise),
and, except for the diagrams with all ladders or all arcs, the
separable diagrams change the  edge-type exactly once (from
arc to ladder).  
When 
$N$ is  comparable to $K_1, \cdots, K_{n-1}$,
the contributions from the separable diagrams are comparable 
to that from the diagram of all edges.

Collecting the terms corresponding to the separable diagrams 
 we have
\bean
&&\mu^2NM\prod_{i=1}^n\sigma_i^2
 \prod_{j=1}^{n-1}K_j\big(\prod_{k=1}^{n-1} K_{k}
 +N\sum_{i=1}^{n-1}K_1\cdots \widehat K_i\cdots K_{n-1}\big)
 \eean
  where $\widehat K_i$ means that $K_i$ is absent in the
 product. Dividing it by $N^2\prod_{i=1}^{n-1}K_i^2$ and accounting for the temporal aspect of transmission as in
 the observations following eq. (\ref{1.3}) we obtain
 the claimed result. 

\subsection{Information rate with  pinholes}
The notion of the effective number of
 TRA elements is useful in estimating the channel capacity
as well as the stability condition in the presence of pinholes
since $\sir$ is  given by $N_{\rm eff}/M$ with $C=2B$. 

As the (spatial) subchannel from TRA to each receiver
is the sum of   $NK_1K_2\cdots K_{n-1}$  
paths which  are not
necessarily independent, the simplest way for realizing
 Gaussian  interference statistic is to assume large degrees of freedom in frequency $B\beta_c^{-1}\gg 1$
so that each spatial subchannel gives rise to
 a sum of  $B\beta_c^{-1}$ roughly i.i.d.
r.v.s. This  works only  for the BBFS channels. 
For the NBFF channels, we assume
the worst-case scenario  $K_{\rm min}=\min{[N,K-1,\cdots, K_{n-1}]}\gg 1$ because each subchannel can be regrouped into a sum of 
$NK_1K_2\cdots K_{n-1}/K_{\rm min}$ terms
each of which is a 
 sum of $K_{\rm min}$ i.i.d.  
r.v.s. 

Due to the normalization (\ref{power}) the input-output relation in (\ref{inout})
 and the discussion in Section~4 ({Case 1 \& 2}) remain valid
if $N$ is replaced by  $N_{\rm eff}$. 
In particular, the same optimal information rate $P/\nu$ is achieved in the regime $N_{\rm eff}\wedge M\gg P/(B\nu)$.

As analyzed before, when the condition 
$N, K_1,..,K_{n-1}\gg 2^n$ is satisfied, $N_{\rm eff}$
is the harmonic sum of $N, K_1, ..., K_{n-1}$ and therefore
we have the estimates: $K_{\rm min}/n\leq N_{\rm eff}
\leq K_{\rm max}/n $ where $K_{\rm min}$ and
 $K_{\rm max}$ are the minimum and maximum of 
$N, K_1, ..., K_{n-1}$, respectively.
On the other hand, when $N, K_1,..,K_{n-1}$ $\ll 2^n$,
diagrammatic  analysis shows that 
$N_{\rm eff}$ diminishes exponentially with
the number of screens, making the alternative  regime  
$N_{\rm eff}\leq P/(B\nu)$ much more likely and resulting in
low information rate $BN_{\rm eff}$ (cf. Case 2, Section~4). In other words, a long chain of independently fluctuating media
separated by a series of screens of {\em sparse} pinholes is detrimental
to time reversal (and perhaps any) communication systems

\section{Conclusions}

We have analyzed the time reversal propagation
in independent-multipath Rayleigh-fading MIMO-channels
with or without pinholes. The focus of the analysis is
the stability condition, the multiplexing gain and the multi-screen
pinholes effect.
The main results  are (i)
that the stability holds when $MC\ll N_{\rm eff}B$ for the
BBFS channels and $M\ll N_{\rm eff}$ for
the NBFF channels
where $N_{\rm eff}$ is the effective number of
TRA-elements, (ii) that 
the optimal  information rate $P/\nu$ under the power and
noise constraints  is achieved
in the regime $N_{\rm eff}\wedge M\gg P/(B\nu)$ and
(iii) that  the effective number of TRA-elements
is asymptotically  
  the harmonic sum  of TRA-elements and the numbers of pinholes on all $n-1$ screens when  the numbers
of TRA-elements and the pinholes of each screen   are 
greater than $2^n$.
The notion of  the effective number of TRA elements 
is introduced for the first time and shown to
be useful in analyzing  stability and capacity in the presence
of pinholes. 



\end{document}